\documentclass[10pt,a4paper]{IEEEtran} 

\usepackage{stfloats}   
\usepackage[pdftex]{graphicx}
\usepackage{color}                
\usepackage{amsmath}          
\usepackage{amsfonts}
\usepackage{mathrsfs}
\usepackage{hyperref}		

\usepackage{framed}

\usepackage[english]{babel}  
\usepackage{algorithmic}      
\usepackage{algorithm}        
\usepackage{hyperref}         
\newcommand{\be}{\begin{equation}}
\newcommand{\ee}{\end{equation}}
\date{today}
\begin{document}
\title{Pilot-wave quantum theory \\in discrete space and time 
\\ and the principle of least action}

\author
{\IEEEauthorblockN{Janusz Gluza,}
\IEEEauthorblockA{Institute of Physics, University of
Silesia, \\ Uniwersytecka 4, PL-40-007 Katowice, Poland, \\
Email: janusz.gluza@us.edu.pl }

{\IEEEauthorblockN{Jerzy Kosek,}
\IEEEauthorblockA{Institute of Physics, University of
Silesia, \\ Uniwersytecka 4, PL-40-007 Katowice, Poland, \\
Email: jkosek@us.edu.pl  }                                             \\
\hfill {\today}
} }

\maketitle

\begin{abstract}
\boldmath 
The  idea of obtaining a pilot-wave quantum theory on a lattice with discrete time is presented.  The motion of quantum particles 
is described by a $|\Psi|^2$-distributed Markov chain. 
Stochastic matrices of the process are found by the discrete version of the least-action principle. 
Probability currents are the consequence of Hamilton's principle and the stochasticity of the Markov process is minimized.  
As an example, stochastic motion of single particles in a double-slit experiment is examined.

\end{abstract}

\section{Introduction\label{Intro}}

In classical physics the motion of a system of particles can be elegantly described by Hamilton's principle of least action. It states that for given initial and final space-time configurations the real path of the system is the one for which an action takes a stationary value.
In quantum mechanics formulated by Heisenberg in 1925 and by Schr\"odinger in 1926, the situation is different. It describes time evolution of the wave function (in the Schr\"odinger picture), {while} the classical notion of a sharp trajectory followed by a physical system is rejected. {One of the alternatives} is a pilot-wave formulation of quantum mechanics  proposed  by de Broglie in 1927 \cite{Broglie27}, and re-discovered  by Bohm in 1952 \cite{Bohm52}, where particles have definite positions during their motion, similarly as in classical mechanics. 
This idea {was later} extended to quantum field theories (QFT), both bosonic and fermionic. In particular, some Bell-type QFTs describe 
creation and annihilation  of particles,  which, in addition, follow real trajectories \cite{Durr_03}, \cite{Durr_04},  \cite{Durr_5B}, \cite{Durr_13}. {These models deal with {\it continuous space-time}, and generalize a lattice quantum field theory proposed by Bell \cite{Bell84}, where the latter is based on specific choices of probability currents and jump rates.}

Despite these prominent achievements there are several motivations for developing an analogical approach to quantum (field) theory in {\it discrete space-time}. 
First, space-time is treated as continuous both by classical and standard quantum physics. However, one cannot exclude discrete space-time hypothesis. The situation can be similar to the discovery of  a discrete nature of such fundamental quantities as quanta of energy, charge or angular momentum, etc. Of particular importance is the development of a quantum theory of gravity. Different proposals based on the idea of discrete space-time were given (in particular, see \cite{Vassallo14}, the proposal focused on loop quantum gravity, which poses elementary extensions of space as the primitive ontology of the theory). Also the notion of a digital universe with discrete space-time is commonly used and exploited now (see, e.g. \cite{Zenil13}). Moreover, discrete space-time models are well suited for calculations, and de facto all computer calculations are discrete in nature. 

A quantum model aimed to develop  a Bell-type stochastic process on a lattice in discrete time was studied in \cite{Tumulka05}. It was proven that a genuine analog of Bell's process does not exist in discrete time, however, proposals of processes that could be used as a replacement were given. The problem is that in the discrete case there is not an obvious formula for the net probability current between discrete states, such that it could substitute the continuous case (see Eq. \ref{B_current} below). 

In this paper we present a new pilot-wave model on a lattice configuration space in discrete time. We take the positions of the particles as the primary variables, similar to Bohmian Mechanics and Bell-type QFTs \cite{Durr_5B}. The theoretical scheme presented here is used to find stochastic paths for massive particles in quantum theory. 

We do not assume any arbitrary formula for a probability current. Rather, we describe the motion of quantum particles by a $|\Psi|^2$-distributed Markov chain, where stochastic matrices for any two subsequent instants are found by Hamilton's principle. Thus, probability currents are the consequence of Hamilton's principle. {\it We extend the principle  to a lattice configuration space.} It states that for given initial and final quantum distributions|specified by a state vector $\Psi$|the action averaged over a statistical ensemble of identical systems takes a minimal value. This allows us to find unique stochastic matrices of the process. Additionally, the stochasticity of the Markov process is minimized.  In the case of single non-relativistic particles it means that 
{\color{black}their mean square displacements} over time are minimized.

{\color{black}Finally, Hamilton's principle on a lattice with discrete time 
can be viewed as an optimal transport problem, where an average action is equivalent to the so-called 
{\color{black} optimal transport cost.}  
(For an introduction to the optimal transport, see  \cite{Salvare08},  \cite{Villani09}).
We suppose that  numerical methods developed in the field of optimal transport could be used (or adopted) in computation of quantum phenomena on a discrete space-time lattice.} 

The plan of this article is as follows. Section~\ref{BohmianM} briefly presents Bohmian mechanics and some other pilot-wave models with continuous time and continuous or discrete space, providing a theoretical background to the discrete space-time model. 
Section~\ref{secMatrix} defines stochastic matrices for a quantum system on a lattice configuration space.
Section~\ref{secHamilton} formulates the principle of least action on a lattice configuration space in discrete time. 
Section~\ref{secProb} defines minimal transition probabilities and proves it to be valid for our model.
Section~\ref{secRelativ} studies stochastic behavior of single massive particles.
Section~\ref{secSlits} presents 
an example of double-slit experiment with massive particles. 
Section~\ref{secConclusions} summarizes our results and gives concluding remarks.
Appendix~\ref{secAlgorithm} presents an algorithm for computing stochastic matrices with the minimal transition probabilities.   
{\color{black}Appendix~\ref{secWasserstein} discusses the link between Hamilton's principle and the optimal transport problem. 
}

\section{ Continuous-time pilot-wave models\label{BohmianM}}
In Bohmian mechanics space-time is continuous, and the state of the system at any time $t$ is described by {\it a configuration} $Q(t)=(\mathbf{Q}_1(t), .., \mathbf{Q}_N(t))$ of $N$  point  particles moving in real space $\mathbb{R}^{3}$. The wave function $\Psi_t(q)$ given by the Schr\"odinger equation 
\be
i\hbar\frac{\partial \Psi_t(q)}{\partial t} = 
\Big(-\sum_{k=1}^N\frac{\hbar^2}{2m_k}
\nabla_k^2+ V(q)\Big)\Psi_t(q)
\label{eq:Schrodinger}
\ee
plays the role of a guiding field for the particles; $V: \mathbb{R}^{3N} \rightarrow \mathbb{R}$ is the potential function; $\nabla_k$ is the gradient relative to the space coordinates of particle $k$, $m_k$ is its mass. The equations of motion are
\be
\frac{d\mathbf{Q}_k}{dt} = \frac{\mathbf{ j}_t^k(q)}{|\Psi_t(q)|^2}
\Bigl|_{q=Q(t)},
\label{e_guidance}
\ee
where 
\be
\mathbf{j}_t^k(q)=\frac{\hbar}{m_k} \text{Im}(\Psi^{\ast}_t(q)\mathbf{\nabla}_k\Psi_t(q))
\label{s_current}
\ee
is the usual quantum current. The system itself, depending upon its initial position, follows a deterministic trajectory. 

The guidance equation, Eq. \ref{e_guidance}, implies the property called {\it equivariance}: a statistical  ensemble of systems having a distribution in positions $|\Psi_{t_0}(q)|^2$ at $t_0$ preserves the character of this distribution at any later time $t$, i.e. the distribution is $|\Psi_t(q)|^2$. As a consequence, predictions of Bohmian mechanics are identical to the predictions of standard quantum mechanics.

Other types of pilot-wave models are stochastic models initiated by Nelson \cite{Nelson66}, \cite{Nelson85}. Now, instead of Bohmian guidance equation one has a Langevin equation with stochastic parameters, while the wave function $\Psi$ still satisfies the Schr\"odinger equation.
Both de Broglie-Bohm and Nelson's models can be generalized in a way that the guidance equation, Eq. \ref{e_guidance}, is supplemented with additional terms~\cite{Deotto98}, \cite{Bacciagaluppi99}. For example, in the case of a single particle system a generalized equation is \cite{Bacciagaluppi99}
\be 
\frac{d\mathbf{Q}}{dt}= \frac{\mathbf{j}_t(q)+\mathbf{j}_t(q)_ \emph{DG}}{|\Psi_t(q)|^2} +  \alpha\frac{\hbar}{m}\frac{\nabla |\Psi_t(q)|^2}{|\Psi_t(q)|^2}+\sqrt{\alpha}\frac{d\omega}{dt}\Bigl|_{q=Q(t)},
\label{e_guidance_general}
\ee
where $\nabla\! \cdot {\mathbf{j}_t(q)_{\emph{DG}}} =0, \alpha$ is a free parameter, and $d\mathbf{\omega}$ is a Wiener process with $\overline{d\mathbf{\omega}}=0$ and $\overline{(d\omega)^2}=\hbar/m$. 

An important feature is that the generalized guidance equation, Eq. \ref{e_guidance_general}, gives rise to the same quantum distributions $|\Psi_t(q)|^2$, and simultaneously the trajectories are different, depending on the choice of the parameters. For $\alpha\neq0$ we get stochastic theories,  while for $\alpha=0$ deterministic ones. 

So there is an infinity of possible wave-particle models in continuous space-time, both deterministic and stochastic. Bohmian mechanics is the one with the usual current $\mathbf{j}$ ($\mathbf{j}_\emph{DG} =0$) and no stochasticity $(\alpha = 0$).

\subsection{Discrete-space and continuous-time models}\label{Bell}

An extension of Bohmian mechanics into the discrete configuration space  and continuous time is Bell's model \cite{Bell84}. It presents a Markov pure jump process $(Q_t)_{t\in\mathbb{R}}$ on a  lattice configuration space $\mathcal{Q}$. Bell aimed to reproduce the quantum mechanical predictions for fermion number density in space. The same method|properly generalized|can be used to find stochastic evolution for any discrete beables (i.e. the quantities supposed as objective elements of reality), both in non-relativistic quantum mechanics and in QFT.
\footnote{In Bell's model, a continuous real  space $\mathbb{R}^3$ is replaced by 3D spatial lattice $\Lambda$. At given time $t$, the actual configuration $Q$ of fermion particles of the world is one of the possible lists of integers $q=(q_1, \ldots, q_N) \in \mathcal{Q}$, where $N$ is the maximal index of the lattice sites, and $q_k$ are eigenvalues of fermion number operators  acting at particular sites $k$ of the lattice $\Lambda$ ($q_k\in\{1,2,\dots,4M$\}, where $M$ is the number of Dirac fields).}  

Dynamics of the actual (field) configuration is stochastic|it is a consequence of the discretization of space. The transition probability from a configuration $q$ to other configuration $q'$ ($q'\ne q$) during a {\it small} interval $dt$ is defined by Bell as 
\be
 \mathbb{P}_t(q \to q')
=\Bigg \{
\begin{tabular}{ccc}
  $J_t(q', q)dt/\mathbb{P}_t(q),$  &  & $ J_t(q', q)>0$,   \\
             $0,$                             &  & $ J_t(q', q)\le 0$, \\
  \end{tabular}
\label{Bell_jumps}
\ee
where $J$ stands for the probability current 
\be 
J_t(q', q)=\frac{2}{\hbar}\text{Im}[\langle\Psi_t|P(q')HP(q)|\Psi_t\rangle], 
\label{B_current}
\ee 
$\Psi_t$ is the state vector of a quantum (field) theory, evolving in a Hilbert space $\mathscr{H}$ according to the Schr\"odinger equation;  $H$ is the Hamiltonian, $P(q)$ is a projection to $\mathscr{H}_q \subseteq \mathscr{H}$, and the $\mathscr{H}_q$ form an orthogonal decomposition, $\mathscr{H} =\bigoplus_{q \in \mathcal{Q}} \mathscr{H}_q$; $\mathbb{P}_t(q)$ is the probability distribution at time~$t$ 
\be
\mathbb{P}_t(q)=\langle\Psi_t|P(q)|\Psi_t\rangle.
\ee
The probability $\mathbb{P}_t(q \to q)$ to stay in the same state $q$ is 
\be
\mathbb{P}_t(q \to q)= 1 - \sum_{q'\ne q} \mathbb{P}_t(q \to q').
\label{Bell_jumps_qq}
\ee
Notice that the current $J$ is defined in analogy to the current $\mathbf{j}$, Eq.~\ref{s_current}. As it is antisymmetric, i.e., $J_t(q, q')=-J_t(q', q)$, thus Eq.~\ref{Bell_jumps} implies $\mathbb{P}_t(q \to q')=0$ or $\mathbb{P}_t(q' \to q)=0$. So at least one of transitions $q\to q'$ or $q'\to q$ is forbidden.

Solution Eq. \ref{Bell_jumps} can be generalized \cite{Bacciagaluppi99}, \cite{Vink93}. 
{\color{black}For example, one can {\it add} to  $ \mathbb{P}_t(q \to q')$ defined in Eq.~\ref{Bell_jumps} any solution $\mathbb{P}^0_t$ of the homogeneous equation
\be
\mathbb{P}^0_t(q \to q')\mathbb{P}_t(q)=  \mathbb{P}^0_t(q' \to q)\mathbb{P}_t(q').
\label{Vink_jumps_qq'}
\ee }
This generalization makes both of the transitions  $q\to q'$ and $q'\to q$ possible. For all 
$\mathbb{P}^0_t(q \to q')=0$ one gets Bell's solution, Eq.~\ref{Bell_jumps}, with  {\it minimal transition probabilities} (or, equivalently, with minimal jump rates). This means that at least one of the transitions $q\to q'$ or $q'\to q$ is forbidden.

\section{Stochastic matrices for quantum systems\label{secMatrix}}

In Sections~\ref{secMatrix} - \ref{secSlits} we develop a new pilot-wave quantum model on a lattice in discrete time. Basic assumptions of the model are as follows: 
{\color{black}We consider a system of $N$ structureless and distinguishable particles and assume that all physical beables are definite positions of the particles. 
The configuration space $\mathcal{Q}$ is an ensemble of configurations of discrete particle positions. The state of the system at time $t$ is described by its actual configuration $Q_t=q \in \mathcal{Q}$. 
The dynamics of the configuration is stochastic - it jumps from its actual position at time $t$ to another position $Q_{t+\tau}=q'\in \mathcal{Q}$ at next time $t+\tau$, where $\tau$ is a discrete time step. Moreover, transition probabilities $\mathbb{P}_t(q\to q')\equiv\mathbb{P}(Q_{t+\tau} = q'| Q_t = q)$, depend only on the actual configuration $Q_t$, and do not depend on the earlier states. This means that an evolution of the state of the configuration 
is a Markov process $(\tilde Q_t)_{t \in \tau\mathbb{Z}}$ on  $\mathcal{Q}$ with discrete time step $\tau$. 

Finally, we assume also that the transition probabilities depend on the state vector $\Psi$ which is the solution to the appropriate Schr\"odinger equation 
defined on $\mathcal{Q}$ 

\be
i\hbar\frac{\partial \Psi_u}{\partial u} = H \Psi_u,
\label{eq:Schrodinger_QFT}
\ee
where $u \in \mathbb{R}$  is the continuous time parameter. 
Some discrete values of $u$ make discrete instants $t$, when  stochastic jumps happen (i.e., $t=[u/\tau]\in \tau\mathbb{Z}$, where $[x]$ means an integer part of $x$).
\footnote{A similar assumption is presented in \cite{Tumulka05}, where a Markov process in discrete time is obtained by restriction to the integer times of Bell's process $(Q_t)_{t\in\mathbb{R}}$ in continuous time.}

 In the case of spinless particles we have an orthonormal basis 
$\{ |q\rangle : q \in \mathcal{Q}\}$ of a Hilbert space labeled by $\mathcal{Q}$. In general, the basis of a Hilbert space is indexed by the configuration $q$, as well as by additional quantum numbers $m$ that are not related to beables, e.g. spin.  Here we follow Bell who has shown in \cite{Bell66} that spinor-valued wave functions fully account for all phenomena involving spin. The same treatment of spin is in Bohmian Mechanics \cite{Goldstein09}.\footnote{
For example, the wave function of a spin-$1/ 2$  particle is a function $\Psi : \mathbb{R}^3 \to \mathbb{C}^2$, and for $N$ such particles it is a function $\Psi : \mathbb{R}^{3N} \to \mathbb{C}^{2N}.$} 
In this way 
positions can be the primary outputs of the theory, while other quantities (momenta, energy, spin etc.) can be deduced from the positions. 

Now, we define a stochastic matrix  such that the quantum probability distribution  at time $t$
\be
\mathbb{P}_t(q )
= \langle\Psi_t|P(q)|\Psi_t\rangle,
\label{Joz1}
\ee 
is transformed into the  probability distribution at time $t+\tau$
\be
\mathbb{P}_{t+\tau}(q )
=\langle\Psi_{t+\tau}|P(q)|\Psi_{t+\tau}\rangle.
\label{Joz2}
\ee
$P(q)=\sum_m|q m\rangle\langle  q m|$ is a projection.  
{\color{black} Namely, the stochastic matrix $\mathbf{P}_t \equiv \left[ \mathbb{P}_t(q \to q') \right]$ includes probabilities of transitions from $q$ to $q'$, such that the following conditions are satisfied}
\be
\mathbb{P}_t(q)=\sum_{q'} \mathbb{P}_t(q  \to  q') \mathbb{P}_t(q),
\label{Joz3}
\ee
and
\be
\mathbb{P}_{t+\tau}(q')=\sum_q \mathbb{P}_t(q  \to  q') \mathbb{P}_t(q).
\label{Joz4}
\ee
These two equations express conservation of probability at particular sites $q\in \mathcal{Q}$ at $t$ and $q'\in \mathcal{Q}$ at $t+\tau$, respectively. 

The relations Eqs. \ref{Joz1} - \ref{Joz4} are general and do not define stochastic matrices uniquely. However, such matrices can be always defined for any quantum system described by a wave function $\Psi_t(q)$. 
One of the possible solutions is
\be
 \mathbb{P}_t(q  \to  q')=\langle\Psi_{t+\tau}|P(q')|\Psi_{t+\tau}\rangle,
\label{Joz5}
\ee
which means that the transition probabilities depend only on the wave function at latter time $t+\tau$. In this solution the probability is not locally conserved, but only on a global scale, as it involves jumps even from distant sites $q$ to a given $q'$. 

In general, there is a lot of freedom to find solutions for this stochastic matrix, when the only requirement is that it restores given probability distributions, Eqs. \ref{Joz1} and \ref{Joz2}. A unique solution locally conserving probability, based on the least action principle, is proposed below.

\section{Hamilton's principle on a lattice in discrete time \label{secHamilton} }

Hamilton's principle allows to formulate classical mechanics in a very general way. The formulation assumes continuous configuration space  and continuous time. Now we adjust the principle to the discrete space-time case.
It involves stochastic processes.

Let us consider a statistical ensemble of $\mathcal{N}$ identically prepared systems undergoing identical initial conditions, which take up sites $q^{(\nu)}\in \mathcal{Q}$ at time $t$, $\nu=1,\ldots, \mathcal{N}$, and jump to sites $q'^{(\nu)}\in \mathcal{Q}$ at $t+\tau$. 

We shall assume that the contribution from a single jump from $q^{(\nu)}$ to $q'^{(\nu)}$ 
is the classical action, i.e., the time integral of the Lagrangian taken along the appropriate classical path $q^{(\nu)}(u), u\in\langle t,t+\tau\rangle$, as if the system really follows this path
\be
S_{t}(q'^{(\nu)},q^{(\nu)}) =\text{min}
 \int\limits_{t}^{t+\tau}L(\dot{q}^{(\nu)}(u),q^{(\nu)}(u))\, du.
\label{Eq:action}
\ee 
Now, let us define a total action for the statistical ensemble
\be
\text{S}_t(\mathcal{N})=\sum\limits_{\nu =1}^{\mathcal{N}}{S_t(q'^{(\nu)},q^{(\nu)})}.
\label{Eq:action_total}
\ee
In the lattice space $\mathcal{Q}$ the numbers $S_t(q'^{(\nu)},q^{(\nu)})$ belong to the discrete set of actions calculated along all possible paths connecting sites $q\in \mathcal{Q}_{t}$ and $q'\in \mathcal{Q}_{t+\tau}$, i.e. $S_t(q'^{(\nu)},q^{(\nu)}) \in \{ S_t(q',q) \}$. 
$\mathcal{Q}_{t}$ and $\mathcal{Q}_{t+\tau}$ are subspaces of  $\mathcal{Q}$ such that the wave function is non-zero there at times $t$ and $t+\tau$, respectively. 
Therefore, for $\mathcal{N}_t(q',q)$ systems passing from $q$ to $q'$ we can write
\be
\text{S}_t(\mathcal{N})=\sum\limits_{q'}\sum\limits_q \mathcal{N}_t(q',q)\,S_t(q',q).
\label{443}
\ee
In the limit $\mathcal{N} \to \infty$ the quantity $\mathcal{N}_t(q',q)/\mathcal{N}$ approaches the total transition probability from $q$ to $q'$, which is the transition probability $\mathbb{P}_t(q  \to  q')$ multiplied by the probability $\mathbb{P}_t(q)$
\be
\lim_{\mathcal{N}\to \infty} \frac{\mathcal{N}_t(q',q)}{\mathcal{N}} = \mathbb{P}_t(q  \to  q')\, \mathbb{P}_t(q).
\ee
Thus, dividing Eq.~\ref{443} by $\mathcal{N}$ and taking a large $\mathcal{N}$ limit, we get an average value of the action
\be
\bar{\text{S}}(\mathbf{P}_t) = \sum\limits_{q'}\sum\limits_q 
\mathbb{P}_t(q  \to  q')\, \mathbb{P}_t(q)\, S_t(q',q),
\label{S_min}
\ee
which depends on the choice of  a stochastic matrix $\mathbf{P}_t$.
Now,  Hamilton's principle may be readily extended to define
a stochastic matrix for a Markov chain: 

\begin{framed}
{\it Among all possible stochastic matrices constrained by Eqs.~\ref{Joz1} - \ref{Joz4},  a real Markov chain is defined by a matrix $\mathbf{P}_t$ 
minimizing the average action}
\be
\bar{\text{S}}(\mathbf{P}_t)=\text{min }\bar{\text{S}}.
\label{S_min_Joz}
\ee   
\end{framed}

Constraints, Eqs.~\ref{Joz1} - \ref{Joz2}, depend on the wave function, so we call them {\it the quantum constraints} (the other ones, Eqs.~\ref{Joz3} - \ref{Joz4}, are also valid for classical stochastic matrices). 

The description of the stochastic process will be completed by stating initial conditions. We assume that at some time $t_0$ the configuration $Q(t_0)$ is chosen randomly with probability distribution 
$|\Psi_{t_0}(q)|^2$.
The construction of the stochastic matrices guarantees that $Q(t)$ has quantum distribution $|\Psi_t(q)|^2$ at subsequent times $t\in \tau\mathbb{Z}$. Non-equilibrium initial distributions are also valid (similarly as in other pilot-wave models, e.g. in Bohmian-type quantum mechanics), and an open question is how these distributions depend on time. However, we do not consider this question in this paper. 

In this way we have completed a definition of a Markov chain, where the only beables are particles' positions. 
Its properties are studied in Sections \ref{secProb} - \ref{secSlits}.

\section{ Minimal transition probabilities \label{secProb}}

The stochastic quantum dynamics defined here imply validity of the usual rules of probability, despite the fact that specific quantum phenomena are restored. 
For example, the probability of finding a system at site $q_M$ at time $t_M$ is expressed by a sum of total transition probabilities over all the mutually exclusive alternative paths which start from positions $q_0$ at time $t_0$ and follow positions $q_1,\,\dots, q_M$ at  $t_1,\, \dots, t_M$, respectively:
\be
\mathbb{P}_{t_M}(q_M)=\!\!\sum\limits_{q_{M-1}}\!\!\!\ldots \sum\limits_{q_0} 
\mathbb{P}_{t_M,\ldots, t_0}(q_M,\ldots, q_0),
\label{path}
\ee
and the multiplication rule is valid|as in any Markov process
\be 
\mathbb{P}_{t_M,\ldots, t_0}(q_M,\ldots, q_0)=
\left\{\!\prod_{j=0}^{M-1}\! \mathbb{P}_{t}(q_j  \to  q_{j+1})\!\right\} \mathbb{P}_{t_0}(q_0).
\label{path-prob}
\ee


{\color{black}On the other hand,  counterintuitive quantum properties, as an interference (see Section \ref{secSlits}), an entanglement, etc., are naturally embodied  in the pilot-wave model. }

A specific property of our model is that stochasticity of the Markov process is minimized.  An analogical property holds in Bell-type models with continuous time \cite{Durr_5}. 
As it was said in Section \ref{BohmianM} this entails that at least one of two transitions $q\to q'$ or $q'\to q$ is forbidden. We will prove that an even more general property holds in our model. (Yet, we deal with discrete space-time.) 

First, let us define as {\it crossing transitions}, a pair of transitions $a \to b'$ and $b \to a'$ such that 
\be
S_t(a', a) < S_t(b', a)  \text { and } S_t(b', b)  < S_t(a', b),
\label{S_of_crossing}
\ee
and
\be 
\mathbb{P}_t(b', a)  \ne 0  \text { and }  \mathbb{P}_t(a', b) \ne 0,
\ee
where $\mathbb{P}_t(b', a)=\mathbb{P}_t(a \to b') \mathbb{P}_t(a)$ and $\mathbb{P}_t(a', b)=\mathbb{P}_t(b \to a') \mathbb{P}_t(b)$ are the total transition probabilities from $a$ to $b'$ and from $b$ to $a'$, respectively. Notice that in general $a'$ and $b'$ can be different from $a$ and $b$ (see Fig. \ref{scheme_crossing_transitions}). It occurs that Hamilton's principle eliminates at least one of two crossing transitions, i.e. $a\to b'$ or $b\to a'$ is forbidden, as
\be
\mathbb{P}_t(b', a) = 0  \text { or }   \mathbb{P}_t(a', b)  = 0. 
\ee

\begin{figure}[t!]
\centering
\includegraphics[width=0.35\textwidth]{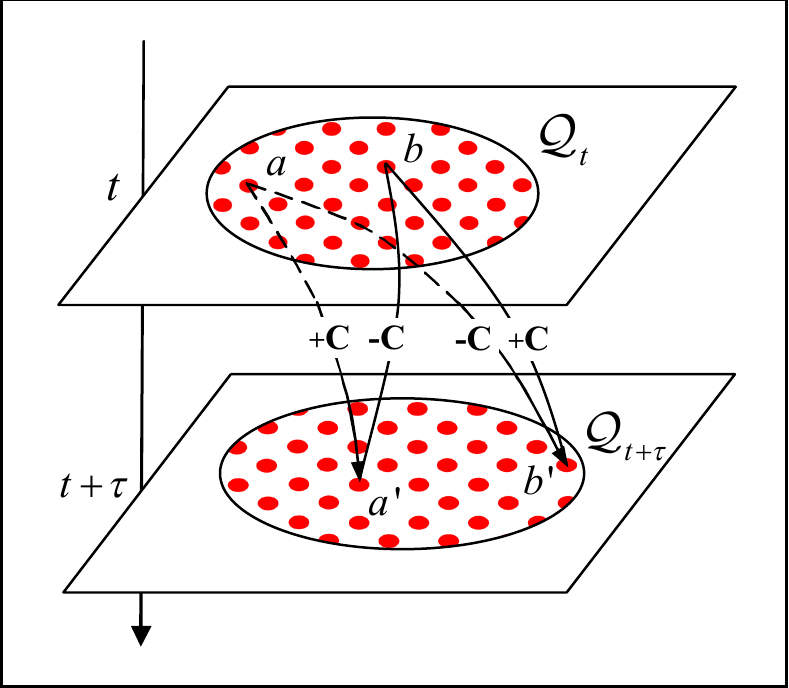}
\caption{A scheme of  transformation which eliminates at least one of superfluous transitions $a\to b'$ or $b\to a'$; $C= \text{min}(\mathbb{P}_t(b', a) , \mathbb{P}_t(a', b))$; $\mathcal{Q}_{t}$ and $\mathcal{Q}_{t+\tau}$ are subspaces of $\mathcal{Q}$ such that  $\Psi$ is nonzero there at times $t$ and $t+\tau$, respectively.}
\label{scheme_crossing_transitions}
\end{figure}

To prove it we assume that proposition is not true, i.e. $\mathbb{P}_t(b', a) \ne 0$ and  $\mathbb{P}_t(a', b)  \ne 0$, while $\bar{\text{S}}(\mathbf{P}_t)$ is minimal. Now we define new total transition probabilities such that they {\it do not change} the probabilities at $a$, $b$, $a'$, and $b'$ (see Fig. \ref{scheme_crossing_transitions}), i.e.
\be
\begin{array}{cll}
\tilde{\mathbb{P}}_t(a', a) =& \mathbb{P}_t(a', a)&\!\!\!\!+\,C,  \cr
\tilde{\mathbb{P}}_t(b', b) =& \mathbb{P}_t(b', b)&\!\!\!\!+\,C,  \cr
\tilde{\mathbb{P}}_t(b', a) = &\mathbb{P}_t(b', a)&\!\!\!\!-\,C,  \cr
\tilde{\mathbb{P}}_t(a', b) = &\mathbb{P}_t(a', b)&\!\!\!\!-\,C,  \cr
\end{array}
\label{transformation}
\ee
where
\be
C= \text{min}( \mathbb{P}_t(b', a) , \mathbb{P}_t(a', b) )
\ee
(in general, $C$ depends on $a, b, a'$, $b'$ and $t$). Eq.~\ref{transformation} implies that $\tilde{\mathbb{P}}_t(b', a) = 0$ or $\tilde{\mathbb{P}}_t(a', b) = 0$. However, the new value $\bar{\text{S}}(\tilde{\mathbf{P}}_t)$ is less than the old one $\bar{\text{S}}(\mathbf{P}_t)$, because 
\be
\bar{\text{S}}(\tilde{\mathbf{P}}_t) -\bar{\text{S}}(\mathbf{P}_t)= C (S_t(a', a) - S_t(b', a) + S_t(b', b) - S_t(a', b))
\ee
is less than $0$ (see Eq. \ref{S_of_crossing}). Thus we got a contradiction with the assumption that $\bar{\text{S}}$ is minimal, q.e.d.

In particular, when $a'=a\equiv q$ and $b'=b\equiv q'$ {\color{black}and Eq.~\ref{S_of_crossing} holds (e.g. for a "free" Lagrangian such as considered in Section~\ref{secRelativ})} one gets the characteristic feature of Bell-type models.

As a result, the crossing transitions are eliminated in the whole net. 
In the case of single particles in flat space-time this means that space-time trajectories cross themselves only in the space lattice nodes. 

The non-crossing property is especially useful in the case of 1D Markov process on a flat space. In that case it is equivalent to Hamilton's principle, Eq. \ref{S_min_Joz}. Thus, searching for a minimum of the average action can be replaced by searching for non-crossing transitions, which is a simpler task
allowing stochastic matrices to be calculated in an efficient way.

\section{Stochasticity of motion  \label{secRelativ}} 

Hamilton's principle implies that the crossing transitions in $\mathcal{Q}$ are eliminated.  In consequence, the stochasticity of the motion in $\mathcal{Q}$ is minimized. Roughly speaking, it is because we get no more distant  jumps than necessary for ensuring the quantum distributions.

To illustrate this, consider the motion of free single particles. We take Lorentz invariant action in Eq. \ref{S_min} \cite{Landau73} 
\be
\begin{array}{cll}
S_t(\mathbf{q}',\mathbf{q})&=&-mc^2\int\limits_t^{t+\tau}\, \sqrt{1 - \dot{\mathbf{q}}^2(u)/c^2} du
\label{S_mass}
\end{array}
\ee
where $m$ is particle's mass. In the non-relativistic limit we get 
\be
S_t(\mathbf{q'},\mathbf{q})=-mc^2 \tau
\left(1-\frac{ |\mathbf{q'}-\mathbf{q}|^2}{2 c^2\tau^2}\right).
\label{S_mass_discrete}
\ee
Now Eq. \ref{S_min} reads
\be
\bar{\text{S}}(\mathbf{P}_t)=K+\frac{m}{2\tau}\sum\limits_{\mathbf{q}'}\sum\limits_{\mathbf{q}} \mathbb{P}_t(\mathbf{q} \to \mathbf{q}')\, \mathbb{P}_t(\mathbf{q})\, 
|\mathbf{q'}-\mathbf{q}|^2,
\label{S_min_el1}
\ee
where 
\be
K=-m c^2 \tau
\ee
and we have used
\be
\sum\limits_{\mathbf{q}'}\sum\limits_{\mathbf{q}} \mathbb{P}_t(\mathbf{q} \to \mathbf{q}')\, \mathbb{P}_t(\mathbf{q})\,=1.
\ee
Constant numbers $K$ and $m/2\tau$ can be omitted because the minimum of
\be
\bar{\text{S}}'(\mathbf{P}_t)=\sum\limits_{\mathbf{q}'}\sum\limits_{\mathbf{q}} \mathbb{P}_t(\mathbf{q} \to \mathbf{q}')\, \mathbb{P}_t(\mathbf{q})\, 
|\mathbf{q'}-\mathbf{q}|^2
\label{S_min_el}
\ee
leads to the same matrix elements $\mathbb{P}_t(\mathbf{q} \to \mathbf{q}')$ as the minimum of $\bar{\text{S}}(\mathbf{P}_t)$.
The above formula can be written as 
\be
\bar{\text{S}}'(\mathbf{P}_t)
=\sum\limits_{\mathbf{q}} 
\mathbb{P}_t(\mathbf{q})\Delta_t^2(\mathbf{q}),
\label{S_variance1}
\ee
where
\be
\Delta_t^2(\mathbf{q})=\sum\limits_{\mathbf{q}'}\mathbb{P}_t(\mathbf{q} \to \mathbf{q}')\, |\mathbf{q}'-\mathbf{q}|^2
\label{S_variance2}
\ee
{\color{black}  is mean square displacement of particles moving} from position $\mathbf{q}$ at $t$ to positions $\mathbf{q}'$ at $t+\tau$.\footnote{{\color{black} Eq. \ref{S_variance2} can be expressed as the sum of the squared average distance of jumps and variance of the positions $\mathbf{q}'$ given $\mathbf{q}$, i.e. $\Delta_t^2(\mathbf{q})=(\mathbf{q} -\!\! <\!\!\mathbf{q}'\!\!>)^2$ 
$+\sum_{\mathbf{q}'}\mathbb{P}_t(\mathbf{q} \to \mathbf{q}')\, (\mathbf{q}' -\!\! <\!\!\mathbf{q}'\!\!>)^2$.}} 
The physical meaning of Hamilton's principle is clear now: 
{\color{black} imposed on Eq.~\ref{S_variance1} the principle confines the motion of particles in a way that 
their mean square displacements over time are minimized.}

{\color{black}The 
total mean square displacement, Eq.~\ref{S_min_el}, } is equivalent to the average action in the non-relativistic limit, and the stochasticity of the motion can be measured by it. 
It is easy to see that in a general case of more complex systems the average action $\bar{\text{S}}(\mathbf{P}_t)$ can provide the measure of  stochasticity.

\section{Double slit experiment\label{secSlits}}
As an example, we analyze typical 
double-slit interference of single particles, where a monochromatic plane incident wave illuminates the diaphragm with  
two slits. Here we consider spinless massive particles (let's call them electrons). 

\subsection{An electron wave function} 

The wave function {behind a single slit} can be calculated with the aid of the Feynman path integrals \cite{FeynmanHibbs}. 
Assuming propagation of the wave with {a constant speed $v_y$ along the $y$-axis perpendicular to the diaphragm one has} 
\be
\Psi (x,t)=\int_{-a/2}^{a/2}K(x,t;x_0, t_0) \Psi(x_0,t_0) dx_0,
\label{Feynman1}
\ee
where $\Psi(x_0,t_0)$ is the field on a slit, and
\be
K(x,t;x_0, t_0)=\sqrt{\frac{m}{2i\pi\hbar(t-t_0)}}\text{exp}\left(i\frac{m(x-x_0)^2}{2\hbar(t-t_0)}\right)
\label{Feynman2}
\ee
is the propagator for a free particle of mass $m$. For a plane incident wave we have $\Psi(x_0,t_0)=\text{const}$. Substituting $t-~t_0=y/v_y$  into Eqs. \ref{Feynman1} and \ref{Feynman2} after some algebra we get the wave function at point $(x, y)$ behind the slit
\be
\Psi(x,y)\sim \{C(u_2)-C(u_1)+i[S(u_2)-S(u_1)]\},
\label{Eq:Fresnel_1}
\ee 
where 
\be 
u_1(x,y)=\sqrt{\frac{2}{\lambda y}}\left(x+\frac a2 \right),\,
u_2(x,y)=\sqrt{\frac{2}{\lambda y}}\left(x- \frac a2 \right), 
\label{Eq:Fresnel1}
\ee
$\lambda= h/m v_y $ is de Broglie wavelength of the electron, and
\be
C(u)=\int\limits_0^u \cos{\left(\frac{\pi}{2} t^2 \right)} dt,\;
S(u)=\int\limits_0^u  \sin{\left(\frac{\pi}{2} t^2 \right)} dt
\label{Eq:Fresnel2}
\ee
are the Fresnel integrals. In the case of two identical slits a superposition of a suitably translated field of the single slit, Eq. \ref{Eq:Fresnel_1}, can be applied
\be
\Psi^{(D)}(x,y)=\Psi(x-d/2,y)+ \Psi(x+d/2,y),
\label{Fresnel_2_electron}
\ee
where $d$ is the separation of the slits, $(d/2, 0)$ and $(-d/2, 0)$ are their centers. 

{\color{black}Notice that  the above equation should be regarded as an approximation to the wave function for the lattice version of the 
Schr\"odinger equation, Eq. \ref{eq:Schrodinger_QFT}.} 
\begin{figure}[!t]
\centering
\includegraphics[width=0.33\textwidth]{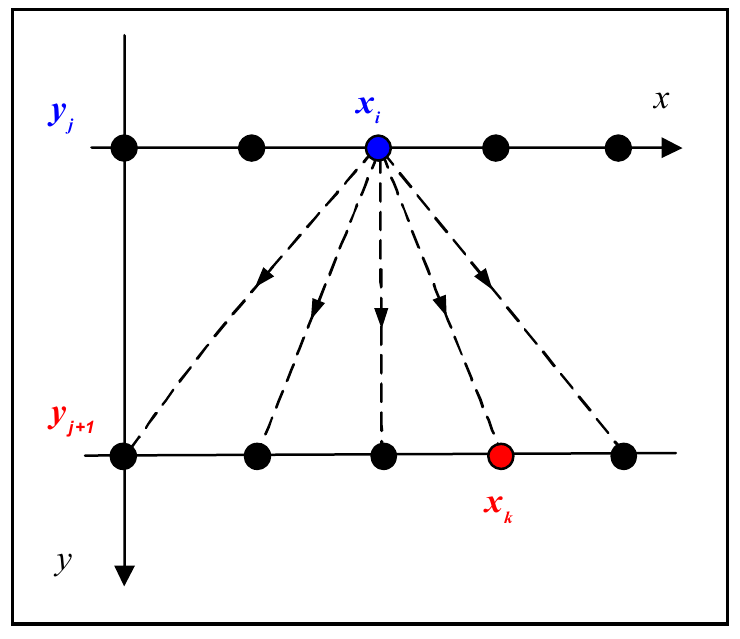}
\caption{A scheme used in the calculation of trajectories: a particle jumps from  position $\mathbf{q}=(x_i, y_j)$ on line $y=\Delta y\cdot j$
at time $t$ to a position $\mathbf{q}'=(x_k, y_{j+1})$ on line $y=\Delta y\cdot (j+1)$ at time $t+\tau$.}
\label{scheme}
\end{figure}

\subsection{Simulation}
The positions {behind} the diaphragm are restricted to the sites of a 2D regular lattice, 
$x_i =\Delta x\cdot i$, $i =0, \ldots ,N_x$, and $y_j =\Delta y\cdot j$, $j =0, \ldots ,N_y$. 
To simplify the calculation, we assume (see Fig. \ref{scheme}):  
\begin{enumerate} 
\item at time $t$ the particle is at site 
$\mathbf{q}=(x_i, y_j)$ on the line $y= \Delta y\cdot j$ parallel to the diaphragm, and 
\item at time $t+\tau$ it jumps randomly
to site $\mathbf{q}'=(x_k, y_{j+1})$, $x_k =\Delta x\cdot k,  k=0, \ldots,N_x$ on the line  $y= \Delta y\cdot(j+1)$; 
$\Delta y = v_y\, \tau$, where $v_y$ is the constant speed of the wave along the $y$-axis. 
\end{enumerate}

Stochastic matrices $\mathbf{P}_t=[\mathbb{P}_t((x_i,y_j)\!\! \to\!\! (x_k,y_{j+1}))]$ are computed  as follows: 
First, the quantum distributions are calculated $\mathbb{P}_t(x_i,y_j)=C_t|\Psi(x_i,y_j)|^2, i=0,\ldots, N_x$ and  $\mathbb{P}_{t+\tau}(x_i,y_{j+1})=C_{t+\tau}|\Psi(x_k,y_{j+1})|^2, k=0,\ldots, N_x$, where $C_t$ and $C_{t+\tau}$ are normalization constants dependent on $\Delta x$.
Then, to find the minimum of Eq. \ref{S_min_el} the imposed (quantum) constraints, Eqs. \ref{Joz1} - \ref{Joz4}, are taken into account. However,  due to computational reasons, directly searching for the minimum would present a challenging task and an equivalent procedure is used based on the non-crossing property of transitions; this made the calculation tractable (see Appendix~\ref{secAlgorithm}, where the algorithm is presented). 

\begin{figure}[!t]
\centering
\includegraphics[width=0.44\textwidth]{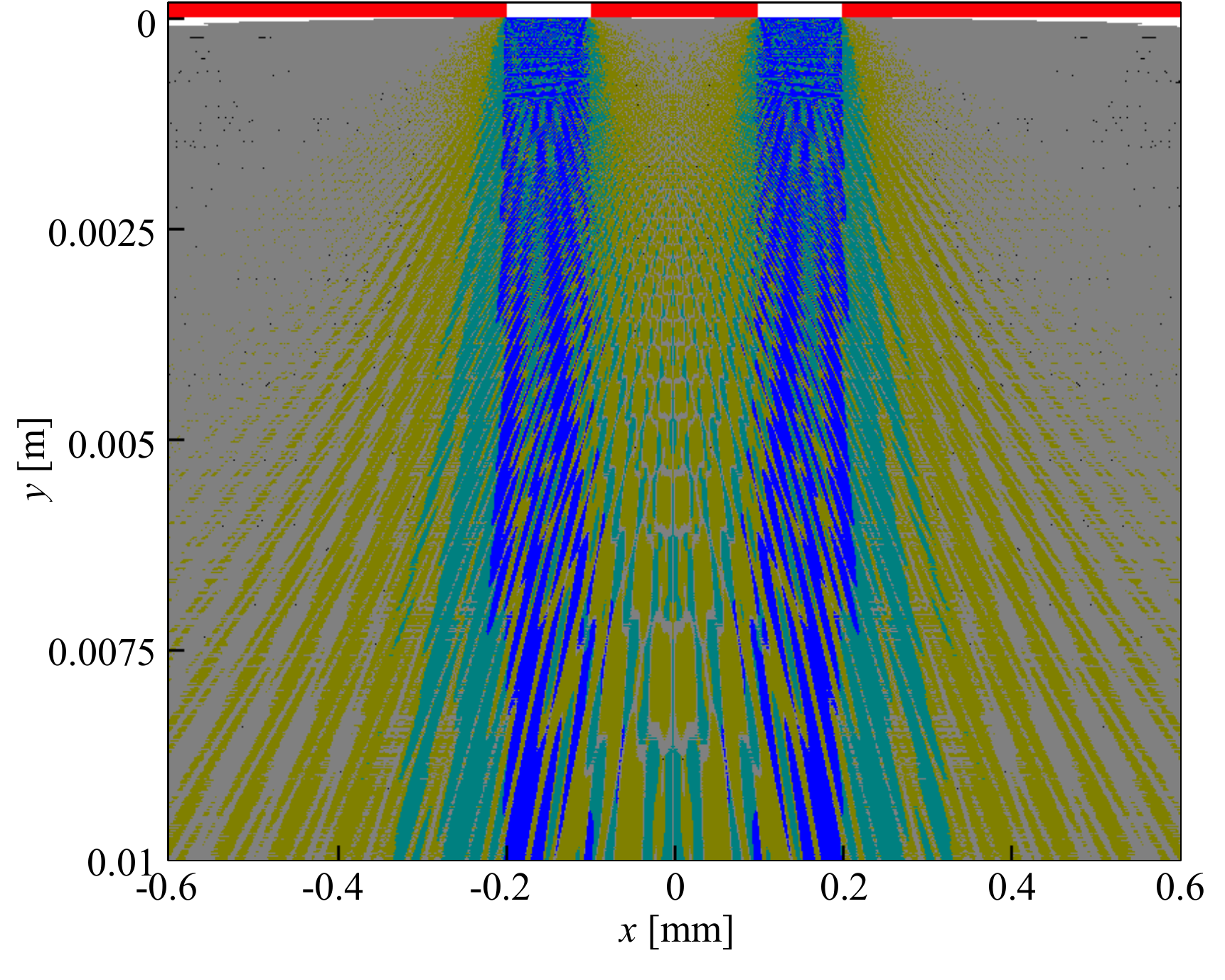}
\caption{A net of  possible transitions in a double slit experiment. The width of the slits $a=0.1$ mm, the distance between them $d=0.3$ mm. The wavelength $\lambda =700$ nm. Colors indicate different ranges of total transition probabilities  related to the maximal one in the net $\mathbb{P}_{max}=0.016$: gray~$[ 10^{-6}, 10^{-3}$), olive~$[10^{-3}, 10^{-2})$, sea green~$[10^{-2}, 10^{-1})$, blue~$[10^{-1},1]$.}
\label{net}
\end{figure}

A net of possible transitions $(x_i,y_j)\!\! \to\!\! (x_k,y_{j+1})$ behind the diaphragm in the double slit experiment is shown in Fig. \ref{net}. 
{Zooming into a picture  we explicitly checked that paths crossed themselves only in the nodes of the net. This means that the stochastic matrix was really minimized.}

The (computer) Monte Carlo algorithm to find the stochastic paths is straightforward. It starts in a diaphragm, where a position of a particle is chosen over the slit/slits width with a uniform distribution  (here it is an equilibrium distribution).
Next, the procedure is repeated recursively, for a given position $\mathbf{q}=(x_i, y_j)$ a pseudo-random number is drawn and, depending on its value and the values of probabilities $\mathbb{P}_t((x_i,y_j)\! \to\! (x_k,y_{j+1})), k=0,\ldots,N_x$, the particle jumps to a new position $\mathbf{q}'=(x_k,y_{j+1})$. 

In Fig. \ref{Fig3} the results are shown. We have taken $\lambda=700$ nm (equal to wavelength of an electron moving with speed $v=10^3$m/s). The band at the bottom shows the distribution of particles on the final screen $y=0.1$ m  in the simulation, where $60\, 000$ particles have been used. For clarity only $0.1\%$ of the total number of trajectories is shown. Two shades of blue are used to visualize individual trajectories in a better way.

The picture reveals the interference 
phenomenon, and we have checked that for an initial equilibrium distribution $|\Psi_{t_0}|^2$ it perfectly restored the equilibrium distributions $|\Psi_t|^2$.  Additionally, the results tended to become closer to the theoretical distributions as more sampling particles were used,  in agreement with the law of large numbers.

What we observe is that starting from the same initial state (source) and passing the diaphragm, particles move along stochastic paths and reach a wide range of positions in the final plane. These trajectories can cross each other. For comparison, in Bohmian mechanics an initial particle's position at the slits implies a single trajectory, and trajectories do not cross each other \cite{Ghose2001, Philippidis1979}.

\begin{figure}[t] 
\centering
\includegraphics[width=0.44\textwidth]{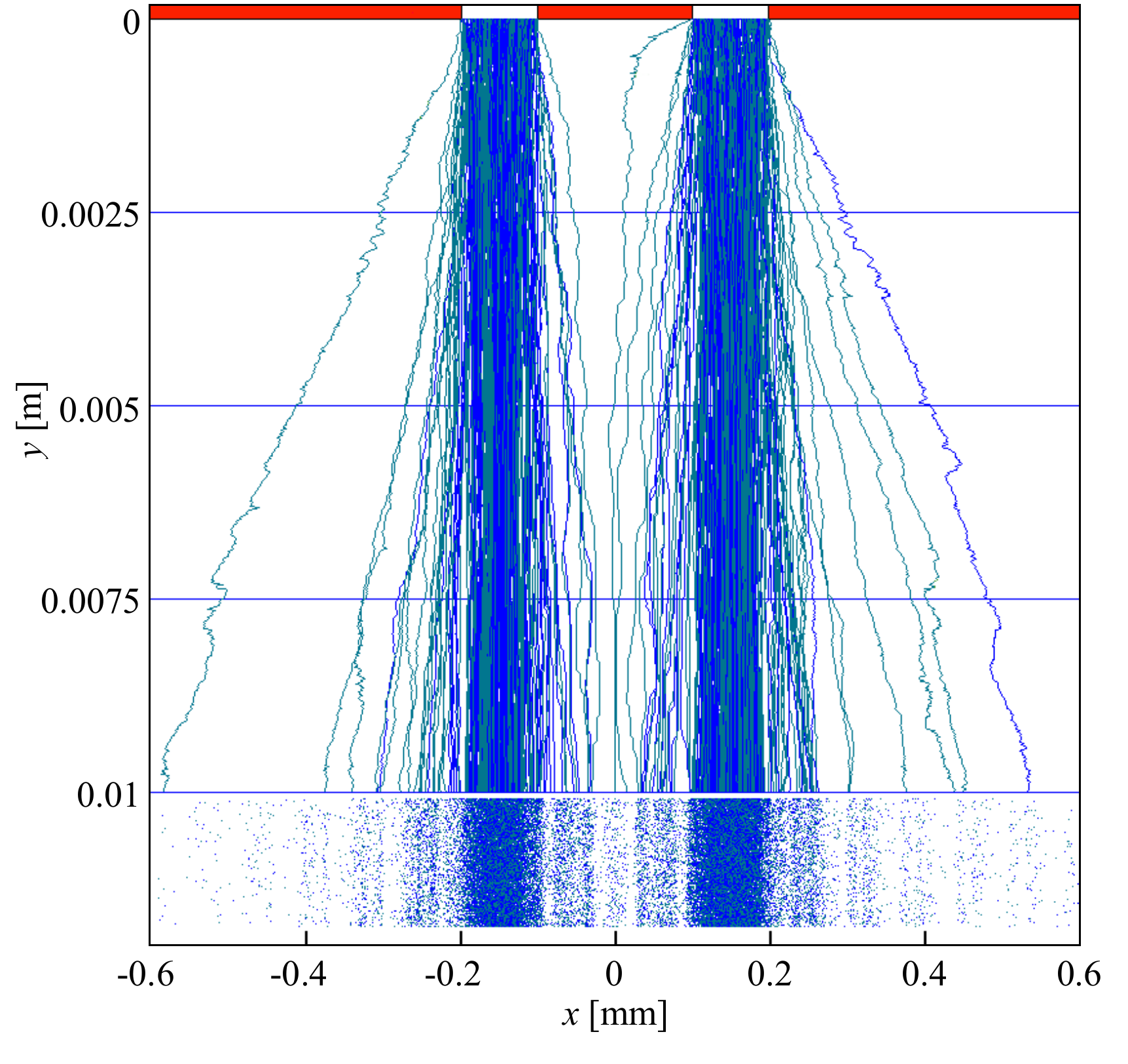} 
\caption{Interference of  single particles in a double-slit experiment, the case of a near field. The width of the slits $a=0.1$ mm, the distance between them $d=0.3$ mm. The wavelength $\lambda =700$ nm.  Theoretical distributions are shown in Fig. \ref{Fig4} for clarity. The interference pattern (Fresnel fringes) shown at the bottom is built up of particles impacts on the screen $y=0.01$~m.}
\label{Fig3}
\end{figure}

\begin{figure}
\centering
\includegraphics[width=0.44\textwidth]{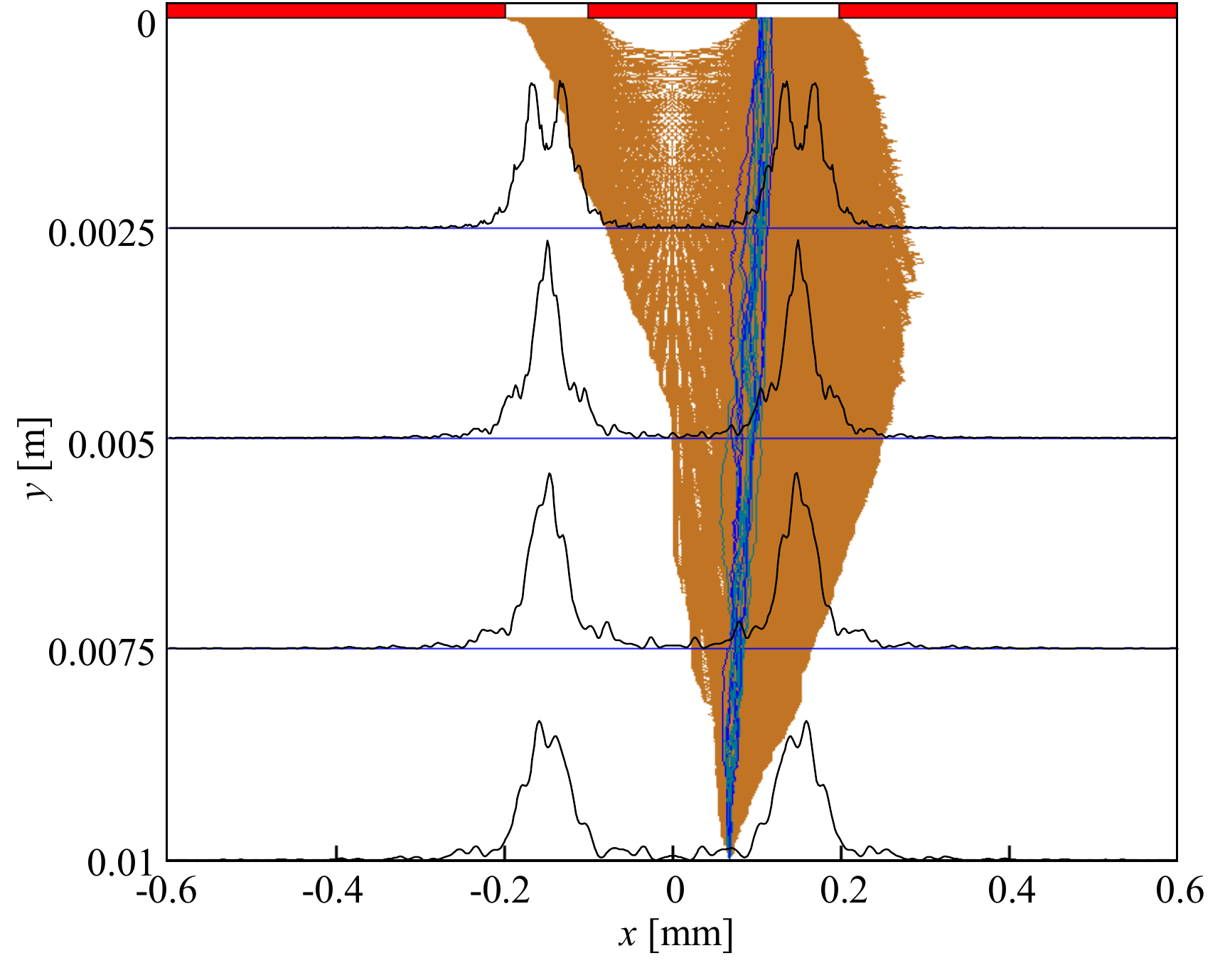}
\caption{Trajectories of  single particles in a simulated double-slit experiment, which ended at a single point on the screen (blue curves), and the net of all other possible trajectories leading to this point (brown color); $a, d, \lambda$ are the same as in Fig \ref{Fig3}. The full black lines are theoretical probability distributions based on Feynman's path integrals. The result explicitly shows that stochasticity of the motion is minimized|although there is a huge number of the possible trajectories the most probable ones form a narrow bunch, while the motion in the rest of the allowed (brown) region is remote.}\label{Fig4}
\end{figure}

Now let us examine the set of all possible paths which start from the same point source and reach the same final state. An example is shown in Fig.~\ref{Fig4}, where a brown region is composed of all these paths. In fact, we draw them going back from the final site on the screen to the source along all different segments of the net with non-zero probabilities $\mathbb{P}_t((x,y) \to (x',y'))$. White areas inside the brown region are not allowed, and for them $\mathbb{P}_t((x,y) \to (x',y'))=0$. Blue curves in Fig.~\ref{Fig4} are the trajectories which ended at the given final state (they were selected from among the whole set of paths get in the simulation). These trajectories form a bunch focused in the neighborhood of the most probable trajectory and make a substantial contribution to the total transition probability (see Eq.~\ref{path}).

Notice, that in general the most probable trajectories are not straight lines. Rather, they are bent curves, attracted into the local maxima or repelled from the local minima of the probability distribution function. A well exposed case is shown in Fig. \ref{Fig8}.  It implies that total action calculated along the most probable trajectories is not a minimal one (in the considered case of a flat space with constant potential, the classical action would be minimal for straight lines). So these curves are not predicted by classical Hamilton's principle of least action. 
 Notice also that Hamilton's principle implies single path for two given space-time points, while here we have a huge number of the possible paths (brown regions).
 
\begin{figure}[tb] 
\centering
\includegraphics[width=0.44\textwidth]{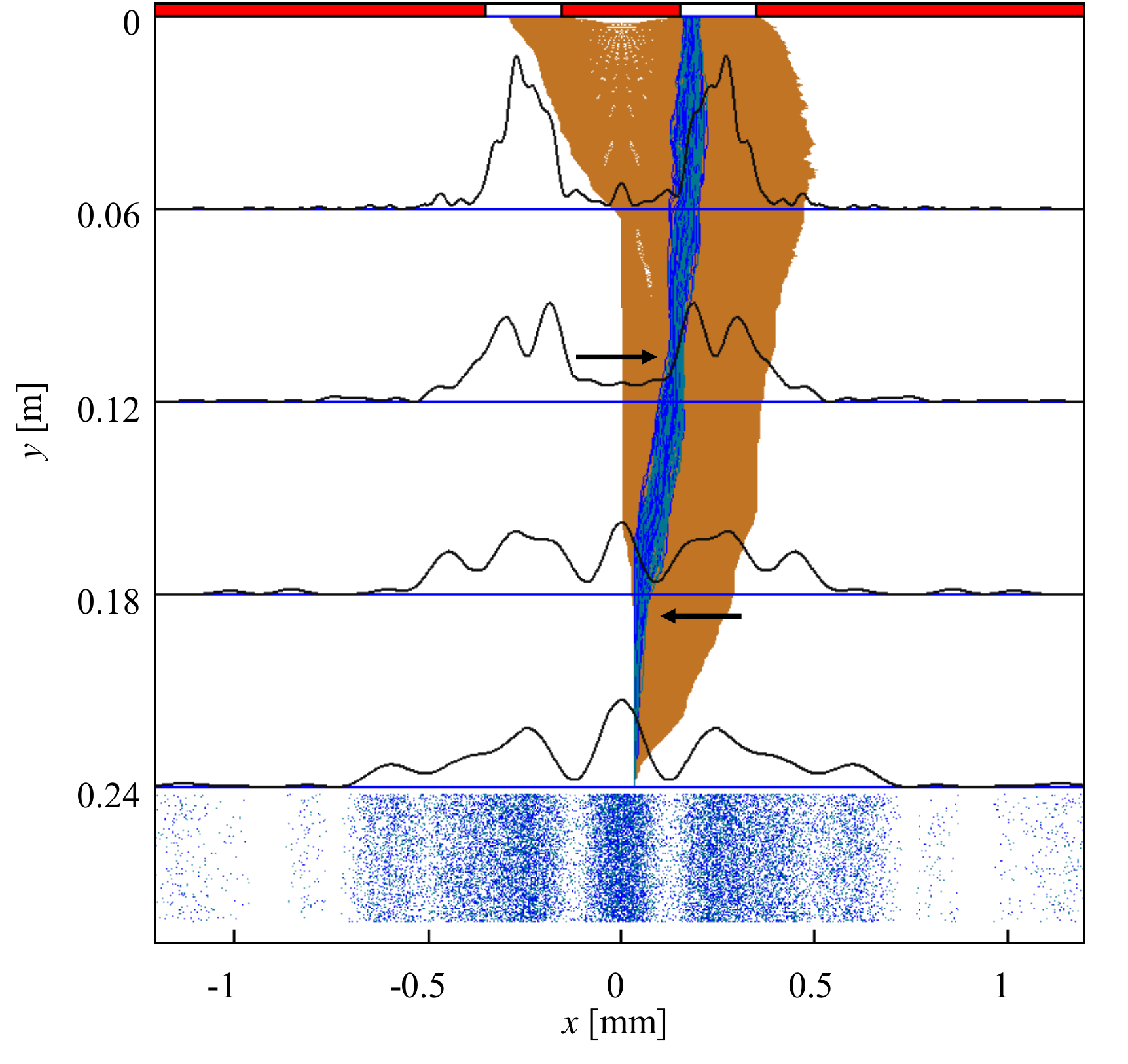}
\caption{Trajectories of  single particles in a simulated double-slit experiment,  which ended at a single point on the screen (blue curves), and the net of all other possible trajectories leading to this point (brown color); the case of a near field. The width of the slits $a=0.2$ mm, the distance between them $d=0.5$ mm. The wavelength $\lambda = 500$ nm. The full black lines are theoretical probability distributions based on Feynman's path integrals. The interference pattern (Fresnel fringes) shown at the bottom is built up of particles impacts on the screen $y=0.24$~m. Black arrows show approximate bending points of the trajectories, attracted into the local maxima of the distribution function. This clearly shows that even the most probable paths are different from Hamilton's classical path of least action.}
\label{Fig8}
\end{figure}

Finally, one can expect convergence to the classical trajectories when quantum effects can be neglected (i.e. when approximation of wave phenomena by 
classical mechanics is valid). Thus, let us study the model in the limit of a short wavelength. For example, for a  wavelength $\lambda=7$ nm we get in a two-slit experiment a set of random trajectories forming two independent parallel bunches (Fig.~\ref{Fig9}). They resemble trajectories of Newtonian (macroscopic) particles which do not possess wave properties and follow straight lines in an empty flat space with constant potential $V$.

\begin{figure}[t]
\centering
\includegraphics[width=0.44\textwidth]{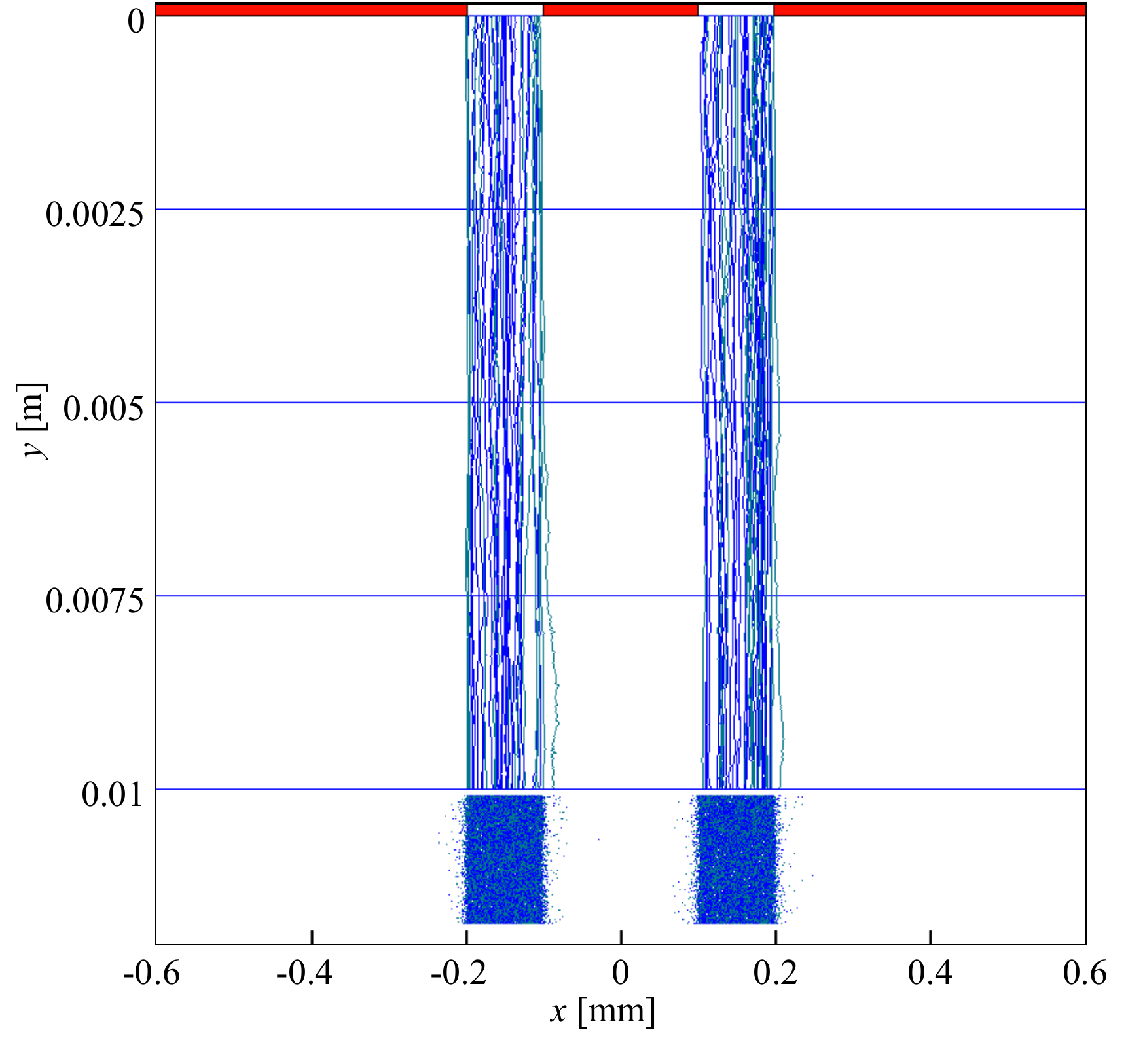}
\caption{Trajectories of  single particles in a double-slit experiment for  the case of a short wavelength $\lambda=7$ nm. The width of the slits is $a=0.1$ mm, the distance between them is $d=0.3$ mm. The trajectories form two narrow bunches as wave properties of particles are negligible, and they behave almost as classical Newtonian particles. Instead of an interference pattern two separated spots appear. 
The pattern shown at the bottom is built up of particles impacts on the screen $y=0.01$ m.}
\label{Fig9}
\end{figure}

\section{Conclusions\label{secConclusions} }

We have shown how to describe the motion of quantum particles by a $|\Psi|^2$-distributed Markov process on a lattice in discrete time.  The discreteness is by itself responsible for the randomness of the motion on the basic level, while Hamilton's principle - extended by us to the discrete case - defines  the stochastic matrices uniquely. An introduction of any additional stochastic parameters does not seem to be justified at that level of description.
The consequence of Hamilton's principle is that the stochasticity of the particles' motion is minimized.

In our approach motion of the system remains random independently of the lattice spacings. Introducing Markovian process is crucial, if time is discretized.
The transformation of a discrete distribution $|\Psi_{t_0}|^2$, e.g. at the slits, into the discrete distribution $|\Psi_t|^2$, e.g. far from the slits, cannot be made in a non-stochastic manner, as it would require transitions from {\it every} (initial or intermediate) state into a single subsequent state. However, such a constraint does not allow  the required transformation for discrete distributions to be constructed as in general  they change in time.

As an outlook, the model sketched here should be extended to the Lorentz invariant form and checked in simulations.
We can take the Lorentz invariant action in calculations of $\bar{\text{S}}(\mathbf{P}_t)$. However, when we take regular space lattice then the stochastic matrices are dependent on that choice, and the model is not Lorentz invariant. The question is if it can be made Lorentz invariant, e.g., assuming that discrete space points are placed randomly according to Poisson statistics, as in causal quantum gravity models (see, e.q. \cite{Henson09} and \cite{Dowker04}). Yet, the idea of a preferred Lorentz frame is also considered (see, e.g., \cite{Valentini97}.


\appendices
\section{An algorithm for computing a minimal stochastic matrix \label{secAlgorithm}}
 
\begin{algorithm}[t]
\caption {For given  initial  $\mathbb{P}_t(x)$ and final $\mathbb{P}_{t+\tau}(x)$ distributions an algorithm computes a `minimal' stochastic matrix $\mathbb{P}_t(x\to x')$.}        
\label{alg1}                           
\begin{algorithmic}
\FOR {$x = x_0$ to $x_N$ }
      	\STATE  $A(x)=\mathbb{P}_t(x)$
		\STATE $B(x) =\mathbb{P}_{t+\tau}(x)$			  
\ENDFOR
\FOR {$x = x_0$ to $x_N$ }
        \FOR {$x' = x_0$ to $x_N$ }
					\IF{$A(x)B(x') > 0$}
									\STATE  $J_t(x',x)$ = min$(A(x), B(x'))$
							 		\STATE $A(x) =A(x) -  J_t(x',x)$
									\STATE $B(x')=B(x') - J_t(x',x)$
									\STATE $\mathbb{P}_t(x\to x')=J_t(x',x)/\mathbb{P}_t(x)$
					\ELSE
									\STATE $\mathbb{P}_t(x\to x')=0$
				   \ENDIF 							  
		\ENDFOR
\ENDFOR
\end{algorithmic}
\end{algorithm}

Let us look at the Algorithm (see the Algorithm 1 frame) for computing a minimal stochastic matrix in the 1D case.  The algorithm finds the minimum of Eq.~\ref{S_min} by searching for non-crossing transitions and it simultaneously encompasses the (quantum) constraints, Eqs. \ref{Joz1} - \ref{Joz4}.  
 Auxiliary numbers $A(x)$ and $B(x)$ used in the pseudo-code are initially equal to the probabilities $\mathbb{P}_t(x)$ and $\mathbb{P}_{t+\tau}(x)$, respectively, $x_0, \cdots, x_N$ are the sites of the lattice. 
$J_t(x',x)$ stands for a probability current (up to a constant factor $\tau$). {\color{black}It is equal to min$(A(x), B(x'))$, i.e. it is the maximal allowed current between sites $x$ and $x'$ at a given step of computation, dependent on the actual values of probability to be transported $A(x)$ from $x$, and an actual value of probability $B(x')$ that can be accepted at $x'$. These values change inside two loops, so }
currents, in general, depend on probabilities at all sites $x$ and $x'$. 

The definition of probability current, done in the algorithm, ensures that at least one of transitions $q\to q'$ and $q'\to q$ is forbidden and $\mathbb{P}_t(q \to q')=0$ or $\mathbb{P}_t(q' \to q)=0$. It resembles prosperities of Bell's definition of the transition probability, discussed in Section~\ref{Bell}. The difference 
is that here the current is not defined in analogy to the standard current, but it is the consequence of Hamilton's principle. 

{\color{black}
\section{Hamilton's principle and the 
optimal transport cost 
\label{secWasserstein}} 
Let $(M, d)$ be a complete, separable metric space (Polish space), and let  $\mu $ and $\nu$ be two probability measures defined on $M$. Then we can define the optimal transport cost  
{\color{black}from $\mu $ to $\nu$} \cite{Salvare08}, \cite{Villani09}. In the case of a discrete metric space it takes the form
\be
C (\mu, \nu)= \inf_{\gamma \in \Gamma (\mu, \nu)} \sum_{q'}\sum_{q} \gamma (q', q)\,c(q', q),
\label{Optimal_cost}
\ee
where  $\Gamma (\mu, \nu)$  denotes the collection of all joint probability distributions $\gamma$  on $M \times M$  with marginals $\mu$  and $\nu$, $c(q',q)$  is the so-called cost function on $M\times M$.

Hamilton's principle, Eqs.~\ref{S_min} and \ref{S_min_Joz}, can be viewed as optimal transport problem. 
$\gamma:  (q',q)\mapsto\mathbb{P}_t(q  \to  q')\, \mathbb{P}_t(q)$  make a collection of joint probability distributions on $\mathcal{Q} \times \mathcal{Q}$ with marginals $\mathbb{P}_t(q)$ and  $\mathbb{P}_{t+\tau}(q')$, and the cost function is associated  with an action $S_t(q',q)$, Eq.~\ref{Eq:action}. 
So the cost function between an initial point $q$ and a final point $q'$ is obtained by minimizing the action among paths that go from $q$ to $q'$.
Finally, an average action $\bar{\text{S}}(\mathbf{P}_t) $ is equal to the optimal transport cost.}

{\color{black}When the cost is defined  in terms of a distance $d$, 
then the optimal transport cost can be replaced by the Wasserstein distance
\be
W_p (\mu, \nu)= \left (\inf_{\gamma \in \Gamma (\mu, \nu)} \sum_{q'}\sum_{q} \gamma (q', q)\,[d(q', q)]^p\right )^{1/p}\!,
\label{Wasserstein_disc}
\ee
where  $p\ge 1$  is the order of the Wasserstein distance. In such a case we get in Eq.~\ref{S_min_el}: the total mean square displacement of particles $\bar{\text{S}}'(\mathbf{P}_t)$, which is equal to the square of the Wasserstein distance of the order $p=2$.


There are several numerical methods proposed in the field of optimal transport problem 
(see, e.q. \cite{Givens84}, \cite{Ruschendorf}, \cite{Pflug}). We suppose that they could be adjusted to simulate quantum-mechanical problems more complicated than presented in this paper, based on Hamilton's principle in discrete space-time.  
}

\section*{Acknowledgment}
{\color{black}
We would like to thank Marek Gluza, Ja\'s Flaten and Piotr Migda\l{} for many discussions and reading of the manuscript.
JK would like to thank his wife Krystyna for her constant support and encouragement during the years of research and preparation of this paper.}



\begin{thebibliography}{10}
\bibitem{Broglie27}L. de Broglie, La nouvelle dynamique des quanta; in:  J. Bordet (Ed.), {\it \'Electrons et photons. Rapports et discussions du cinqui\`eme Conseil de physique tenu \`a Bruxelles du 24 au 29 octobre 1927 sous les auspices de l'Institut international de physique Solvay} (Gauthier-Villars, Paris, 1928). 
English translation is included in: G. Bacciagalluppi and A. Valentini, {\it Quantum Theory at the Crossroads: Reconsidering the 1927 Solvay Conference} (Cambridge University Press, Cambridge, 2009).

\bibitem{Bohm52} D. Bohm, A Suggested Interpretation of the Quantum Theory in Terms of "Hidden" Variables, I and II, Phys. Rev. {\bf 85}, 166 (1952). 

\bibitem{Durr_03} D. D\"urr, S. Goldstein, R. Tumulka, and N. Zangh\`i, {\color{black}Trajectories and particle creation and annihilation in quantum field theory,} J. Phys. A: Math. Gen. {\bf 36}, 4143 (2003). 

\bibitem{Durr_04}D. D\"urr, S. Goldstein, R. Tumulka, and N. Zangh\`i, {\color{black}Bohmian Mechanics and Quantum Field Theory,} 
Phys. Rev. Lett. {\bf 93}, 090402 (2004).

\bibitem{Durr_5B}D. D\"urr, S. Goldstein, R. Tumulka, and N. Zangh\`i, {\color{black}Bell-type quantum field theories,} J. Phys. A: Math. Gen. {\bf 38} (4), R1 (2005). 

\bibitem{Durr_13}D. D\"urr, S. Goldstein, and N. Zangh\`i, {\it Quantum physics without quantum philosophy} ( Springer-Verlag, Berlin, 2013).
\bibitem{Bell84}J. S. Bell, {\color{black}Beables for quantum field theory,} CERN preprint CERN-TH. 4035/84; reprinted in \cite{Bell04}.

\bibitem{Bell04}J. S. Bell, {\it Speakable and unspeakable in quantum mechanics} (Cambridge University Press, Cambridge, 2004).

\bibitem{Vassallo14} A. Vassallo, and M. Esfeld, {\color{black}A proposal for a Bohmian ontology of quantum gravity,} Foundations of Physics {\bf 44} (1) 1 (2014). 

\bibitem{Zenil13} {\it A Computable Universe: Understanding and Exploring Nature as Computation},  H. Zenil (Ed.), (World Scientific, 2013).

\bibitem{Tumulka05} J. Barrett, M. Leifer, and R. Tumulka, {\color{black}Bell's jump process in discrete time}, Europhys. Lett. {\bf 72} (5), 685 (2005).


\bibitem{Salvare08} L. Ambrosio, N. Gigli, and G. Savare, {\it Gradient flows in metric spaces and in the spaces of probability measures}, (Springer Science \& Business Media, 2008).

\bibitem{Villani09} 
C.~Villani, {\it Optimal transport: Old and new}, vol. {\bf  338} (Springer, Verlag Berlin Heidelberg, 2009).


\bibitem{Nelson66} E. Nelson, {\color{black}Derivation of the Schr\"odinger equation from Newtonian mechanics,} 
Phys. Rev. {\bf 150}, 1079 (1966). 

\bibitem{Nelson85} E. Nelson, {\it Quantum Fluctuations} (Princeton University Press, Dordrecht, 1985).

\bibitem{Deotto98}E. Deotto and G.C. Ghirardi, {\color{black}Bohmian mechanics revisited,} Found. Phys. {\bf 28}, 1 (1998). 

\bibitem{Bacciagaluppi99}G. Bacciagaluppi, {\color{black}Nelsonian Mechanics Revisited}, 
Found. of Phys. Lett. {\bf 12}, 1 (1999). 

\bibitem{Vink93} J. C. Vink, {\color{black}Quantum mechanics in terms of discrete beables}, Phys. Rev. A {\bf 48} (3), 1808 (1993). 

\bibitem{Bell66} J. S. Bell, {\color{black}On the problem of hidden variables in quantum mechanics,}  Rev. Mod. Phys. {\bf 38} (3), 447 (1966); reprinted in \cite{Bell04}.

\bibitem{Goldstein09}S. Goldstein, R. Tumulka, and N. Zangh\`i, {\it Bohmian Trajectories as the Foundation of Quantum Mechanics}, in: P. K. Chattaray (Ed.) {\it Quantum Trajectories}, (Taylor \& Francis, Boca Raton, 2010); arXiv:0912.2666 [quant-ph].

\bibitem{Durr_5} D. D\"urr, S. Goldstein, R. Tumulka, and N. Zangh\`i, {\color{black}Quantum Hamiltonians and Stochastic Jumps,} Commun. Math. Phys. {\bf 254} 129 (2005). 

\bibitem{Landau73} L. D. Landau and E. M. Lifshitz, {\it The Classical Theory of Fields} (Pergamon Press, Oxford, 1973).

\bibitem{FeynmanHibbs} R. P. Feynman and A. R. Hibbs, {\it Quantum Mechanics and Path Integrals} (McGraw-Hill, New York, 1965).

\bibitem{Ghose2001}P. Ghose, A. S. Majumdar, S. Guha, and J. Sau, {\color{black}Bohmian trajectories for photons,} Phys. Lett. A {\bf 290} (5), 205 (2001). 

\bibitem{Philippidis1979} C. Philippidis, C. Dewdney, and B. J. Hiley, {\color{black}Quantum interference and the quantum potential,} Il Nuovo Cimento B, Series 11 {\bf 52} (1), 15 (1979). 

\bibitem{Henson09}J. Henson, The causal set approach to quantum gravity; in: D. Oriti (Ed.), {\it Approaches to Quantum Gravity: Towards a New Understanding of Space, Time and Matter,} 393 (Cambridge University Press, New York, 2009). 

\bibitem{Dowker04} F. Dowker, J. Henson, and R. Sorkin, {\color{black}Quantum Gravity Phenomenology, Lorentz Invariance and Discreteness,} Mod. Phys. Lett. A, {\bf 19} (24),  1829 (2004). 

\bibitem{Valentini97} A. Valentini, On Galilean and Lorentz invariance in pilot-wave dynamics,  Phys. Lett. A {\bf 228} (4), 215 (1997).

\bibitem{Givens84}C. R. Givens and R. M. Shortt, A class of Wasserstein metrics for probability distributions, Michigan Math. J., {\bf 31} (2), 231 (1984). 
\bibitem{Ruschendorf}L. R\"uschendorf, The Wasserstein distance and approximation theorems, Z. Wahrsch. Verw. Gebiete {\bf 70} (1), 117 (1985). 

\bibitem{Pflug}G.Ch. Pflug, and A. Pichler, Approximations for Probability Distributions and Stochastic Optimization Problems,  International Series in Operations Research \& Management Science {\bf 163}, 343 (Springer,  New York, 2011).

 







\end{thebibliography}

\end{document}